\documentclass[%
 aip,
 jmp,%
 amsmath,amssymb,
preprint,%
]{revtex4-1}

\usepackage{graphicx}% Include figure files
\usepackage{dcolumn}% Align table columns on decimal point
\usepackage{bm}% bold math
\begin{document}

%\preprint{AIP/123-QED}

\title[]{Addition of SU(3) generators and its Singlet Hilbert space}% Force line breaks with \\
%\thanks{Footnote to title of article.}

\author{Ramesh Anishetty}
\email{ramesha@imsc.res.in}
% \altaffiliation[]{The Institute of Mathematical Sciences}%Lines break automatically or can be forced with \\
\author{T P Sreeraj}%
 \email{sreeraj.tp@gmail.com}
\affiliation{ 
The Institute of Mathematical Sciences, C. I. T. campus, Taramani, Chennai%\\This line break forced with \textbackslash\textbackslash
}%

% \homepage{http://www.Second.institution.edu/~Charlie.Author.}
%\affiliation{%
%Second institution and/or address%\\This line break forced% with \\
%}%

\date{\today}% It is always \today, today,

\begin{abstract}
	We construct the singlet Hilbert space associated with addition of SU(3) generators. This corresponds to the solution of Gauss law in lattice QCD. The normalized basis states are explicitly constructed using Schwinger Bosons. Further, we compute the action of basic singlet operators on these basis states.
%We construct the general irreducible singlet representation extracted from addition of three SU(3) algebra using schwinger boson techniques. The basis states of the representation space are parametrized by the eigenvalues of SU(3) invariant 'casimir' operators. We also calculate the action of the basic SU(3) invariant operators on the basis states.
\end{abstract}

\maketitle

%\begin{quotation}
%The ``lead paragraph'' is encapsulated with the \LaTeX\ 
%\verb+quotation+ environment and is formatted as a single paragraph before the first section heading. 
%(The \verb+quotation+ environment reverts to its usual meaning after the first sectioning command.) 
%Note that numbered references are allowed in the lead paragraph.
%%
%The lead paragraph will only be found in an article being prepared for the journal \textit{Chaos}.
%\end{quotation}

\section{Introduction}
In the Hamiltonian formulation of lattice gauge theory it was envisaged\cite{rs1}  to construct explicit gauge invariant Hilbert space by a simple technique of solving the Gauss law called 'splitting the point' . In this method, the most primitive element of this construction is the addition of two generators  in the adjoint representation $E_1^a,E_2^a$ to give a third $E_3^a$. The Gauss law amounts to looking at only those states which satisfy the condition $(E_1^a+E_2^a+E_3^a)|\rangle=0$. Such states are indeed singlets and it is important to construct these explicitly at each vertex. These states have been explicitly constructed\cite{rs1} for SU(2). Here we pursue the same endeavor for SU(3) which is directly relevant to lattice\cite{lgt} QCD. Furthermore, we construct various basic singlet operators and their action on the above basis states. 

We would like to remark about Clebsch- Gordon coefficients vis a vis the present problem. In the classic addition of generators of algebra we have  the celebrated Wigner-Eckart  theorem which states that the matrix elements of any arbitrary operator with definite irreducible representation (irrep) can be factorized into a Clebsch Gordon term which depends on the azimuthal quantum numbers and a 'reduced matrix element' which only depends upon the casimir quantum numbers of the bra and ket states along with the casimirs of the operator irrep. Clebsch- Gordon coefficients for SU(3) have been explicitly calculated by Prakash et. al \cite{sarat}. Here we are looking at a class of reduced matrix elements which occur in lattice gauge theory. 

The plan of the paper is as follows. In section \ref{su2}, following a short description of the representation of SU(2) in terms of Schwinger bosons, we use them to construct an orthonormal basis for the most general singlet representation from three arbitrary SU(2) irreps. We then calculate the action of basic invariant operators construct out of three SU(2) Schwinger bosons on this basis. 
%We then analyse the invariant space constructed by writing one of the SU(3) in terms of a fermionic representation. 
In section \ref{su3}, we repeat the same for SU(3). 
\section{SU(2)}
\label{su2}
In this section, we will discuss the problem within the simpler context of SU(2).  Consider the direct product space of three arbitrary SU(2) irrep with a basis $|j_1,m_1\rangle|j_2,m_2\rangle|j_3,m_3\rangle$. These basis states are eigenstates of $E_1^2,E_2^2,E_3^2,E_1^{(3)},E_2^{(3)},E_3^{(3)}$ with eigenvalues $j_1(j_1+1),j_2(j_2+1),j_3(j_3+1), m_1,m_2,m_3$ respectively, where $E_1^a,E_2^a,E_3^a$ are the generators of the three SU(2). In order to add these three angular momenta one writes the direct product space as a direct sum of irreducible representation spaces of SU(2) by changing the basis to simultaneous eigenvectors of $(E_1+E_2+E_3)^2, (E_1+E_2+E_3)^{(3)},(E_1+E_2)^2,E_1^2,E_2^2,E_3^2$ with eigenvalues $j_{123},m_{123},j_{12},j_1,j_2,j_3$ (There are two other choices where $(E_1+E_2)^2$ is replaced by $(E_1+E_3)^2$ or $(E_2+E_3)^2$). 
We are interested in extracting a subspace which satisfies the constraint $E_1^a+E_2^a+E_3^a=0$ i.e, a subspace which transform as a singlet under SU(2). As $j_{123}=m_{123}=0$ and $j_{12}=j_3 $, such a space is characterised by three quantum numbers $j_1,j_2,j_3$ only. In order to construct such a singlet subspace it is convenient to write down \cite{prep,schwinger} each SU(2) generators in terms of harmonic oscillator doublets called Schwinger bosons:
%which transform under the fundamental representation of SU(2).
%\subsection{Schwinger boson representation}
%\label{su2sb}
%SU(2) generators can be written in terms of  called Schwinger bosons as follows :
\begin{align}
E^{a} &\equiv 
a^{\dagger}_\alpha\left(\frac{\sigma^{a}}{2}\right)_{\alpha \beta}a_{\beta} \hspace{1 cm} ; ~\alpha, \beta = 1,2.
\end{align}
where $a^\dagger_\alpha, a_\beta$ are the creation, annihilation operator doublets of harmonic oscillators satisfying $[a_\alpha,a^\dagger_\beta]=\delta_{\alpha\beta}$.
The basis of a general, irreducible representation space of SU(2) can be created as 
\begin{align}
|j,m\rangle\equiv |n_1,n_2\rangle= \frac{(a^\dagger_1)^{n_1}(a^\dagger_2)^{n_2}}{\sqrt{n_1!n_2!}}|0\rangle
\label{su2basis}
\end{align} 
Above, $j=\frac{n_1+n_2}{2},m=\frac{n_1-n_2}{2}$ are the eigenvalues of $E^2,E^{(3)}$ and $|0\rangle$ is the harmonic oscillator vacuum $a_\alpha|0\rangle=0$.
%\begin{figure}
%	\includegraphics{3vertex}
%	\caption{The 3 lines meeting at the  3-vertex represents three SU(2) representations written in the Schwinger boson representation}
%	\label{3vertex}
%\end{figure}
$a^\dagger_\alpha$ transform under the fundamental representation of SU(2).  
%In Schwinger bosons representation, the SU(2) generators are broken down into simpler Schwinger bosons which transform under the fundamental representation. 
Therefore, one can directly write down a basis of such a singlet space using Schwinger bosons \cite{rs1} as follows: 
% Basis of such a singlet space is given by:
\begin{align}
|n_1,n_2,n_3\rangle= |l_{12},l_{23},l_{31}\rangle= \frac{(a^\dagger_1\cdot \tilde{a}^\dagger_2)^{l_{12}}(a^\dagger_2\cdot \tilde{a}^\dagger_3)^{l_{23}}(a^\dagger_3\cdot \tilde{a}^\dagger_1)^{l_{31}}}{\sqrt{l_{12}!l_{23}!l_{31}!(l_{12}+l_{23}+l_{31}+1)!}} |0\rangle
\end{align}
where $\tilde{a}^\dagger_{i,\alpha}=\epsilon_{\alpha \beta}a^\dagger_{i,\beta}$ and $n_1=l_{12}+l_{31}, n_2=l_{12}+l_{23},n_3=l_{23}+l_{31}$ are the eigenvalues of the number operators of the three schwinger bosons under consideration. These quantum numbers are related to the original $j_i$ quantum numbers as $n_i=2j_i$. 
The set of basic invariant operators can be constructed by contracting  the operators in the set $\{a^\dagger_i,\tilde{a}_i\}$  with those in $\{a_j,\tilde{a}^\dagger_j\}$.
% in the first row with the second row of table 
%\begin{table}
%\begin{tabular}{|c|c|c|}
%	\hline
%	$2$& $a^\dagger_i$&$\tilde{a}_i$\\
%	\hline
%	$(2^{-1})^T$&$a_i$&$\tilde{a}^\dagger_i$\\
%	\hline
%\end{tabular}
%\label{22star}
%\end{table}
 Action of various invariant operators so constructed on this basis are tabulated in table \ref{su2action}.
\begin{table*}[]
	\begin{center}
		
		\begin{tabular}{ c} 
			\hline 
			\\
			$N_i ~|n_1,n_{2},n_3\rangle = n_i~ |n_1,n_{2},n_3\rangle$\\
			$\big(a^\dagger_3\cdot\tilde{a}^\dagger_i\big)~|n_i,n_{\bar i},n_3\rangle=\frac{1}{2}\sqrt{(m+n_i+n_{\bar i}+4)(m+n_i-n_{\bar i}+2)} ~|n_i+1,n_{\bar i},n_3+1\rangle$\\
			$\big(a^\dagger_3\cdot{a}_1\big)~|n_1,n_{ 2},n_3\rangle=\frac{1}{2}\sqrt{(n_3-n_1+n_{2}+2)(n_1-n_3+n_{2})} ~|n_1-1,n_{2},n_3+1\rangle$\\
%			$\big(a_3\cdot{a}^\dagger_1\big)~|n_1,n_{2},n_3\rangle=\frac{1}{2}\sqrt{(n_1-n_3+n_{2}+2)(n_3-n_1+n_{2})} ~|n_1-1,n_{2},n_3+1\rangle$\\
			$\big({a}_3\cdot \tilde{a}_1\big)~|n_1,n_{2},n_3\rangle=\frac{1}{2}\sqrt{(n_3+n_1+n_{2}+2)(n_3+n_1-n_{2})} ~|n_1-1,n_{2},n_3-1\rangle$\\
			\\
			\hline
		\end{tabular}
	\end{center}
	\caption{Action of various SU(2) invariant operators on the $|n_1,n_2,n_3 \rangle$ basis.} 
	\label{su2action} 
\end{table*}

%\subsection{Fermionic representation}
%\label{su2ferm}
%\begin{figure}
%	\includegraphics{vertex2}
%	\caption{The 3vertex after replacing one of the Schwinger bosons by a fermionic operator}
%	\label{3vertex}
%\end{figure}
%A fermionic representation of the SU(2) generators can be constructed 
%%in a two dimensional irreducible representation space
% as follows 
%\begin{align}
%E^a = q^\dagger_\alpha \Big(\frac{\sigma}{2}\Big)_{\alpha\beta}q_\beta
%\end{align}
%Where $q,q^\dagger$ are fermionic creation annihilation doublet operators satisfying $\{q_\alpha,q^\dagger_\beta\}=\delta_{\alpha\beta}; \{q_\alpha,q_{\beta}\}=0; \{q^\dagger_\alpha,q^\dagger_{\beta}\}=0 $.
% Fermionic creation operators create only the fundamental representation of SU(2).   \begin{align}
%	|j=\frac{1}{2},m=\frac{1}{2}\rangle=|n_{q1}=1,n_{q2}=0\rangle = q^\dagger_1 |0\rangle \hspace{1cm}|j=\frac{1}{2},m=-\frac{1}{2}\rangle= |n_{q1}=0,n_{q2}=1\rangle=q^\dagger_2|0\rangle
%	\end{align}
%	Above, $n_{q1},n_{q2}$ are the eigenvalues of the number operators $N_{q1}=q^\dagger_1\cdot q_1,N_{q1}=q^\dagger_2\cdot q_2 $. These number operators are related to $E^2,E^{(3)}$ as $E^2=3\Big(\frac{N_q}{2}\Big)\Big(1-\frac{N_q}{2}\Big),E^{(3)}=N_{q1}-N_{q2}$ where $N_q=N_{q1}+N_{q2}.$
	
\section{SU(3)}
\label{su3}
%\subsection{Schwinger boson representation}
%\label{su3sb}
A general unitary irreducible representation of SU(3) is labelled by two nonnegative integers (p,q) and is of dimension $\frac{1}{2}(p+1)(q+1)(p+q+2)$. A basis of such a representation space can be created by diagonalizing a complete set of commuting operators given by $C_1=E^aE^a, C_2=d_{abc}E^aE^bE^c, I^2(E)=E^{(1)}E^{(1)}+E^{(2)}E^{(2)}+E^{(3)}E^{(3)},E^{(3)},Y$; where $C_1,C_2$ are the two casimirs and $E^a,I^a,Y$ are the generators of SU(3), Isospin (SU(2) subgroup) and hypercharge (U(1) subgroup) respectively. $d_{abc}$ are the symmetric coefficients defined by  $\{\frac{\lambda_a}{2},\frac{\lambda_b}{2}\}=\frac{1}{3} \delta_{ab}+ d_{abc}\frac{\lambda_c}{2}$, where $\lambda_a$ are the generators of SU(3) in the 3 representation. Such an eigenbasis is denoted as $|p,q,I,M,Y\rangle$ with the eigenvalues of $C_1,C_2,I^2,E^{(3)},Y$ given by $\frac{1}{3}(p^2+q^2+3p+3q+pq),\frac{1}{18}(p-q)(3+p+2q)(3+q+2p),I(I+1),M,Y$ respectively. 
%While adding two SU(3) 'angular momenta', one is interested in splitting the direct product space spanned by $|p_1,q_1,I_1,M_1,Y_1\rangle|p_2,q_2,I_2,M_2,Y_2\rangle$ into direct sum of irreps. This is equivalent to diagonalizing \cite{su3b,mc} $C_1(E_1+E_2), C_2(E_1+E_2),I^2(E_1+E_2),(E(1)+E(2))^{(3)},Y(1)+Y(2),C_1(1),C_2(1), C_1(2),C_2(2), \Gamma$. Unlike SU(2), the same irrep gets repeated \cite{mc} in the  SU(3)  Clebsch Gordon series many times. An additional quantum number namely the eigenvalue of $\Gamma$ is needed to distinguish between such 'multiplicities'. On adding three SU(3)s one requires 4 additional quantum numbers. 
We are interested in the subspace of the direct product space of three SU(3) irreps which satisfies the 8 constraints $E^a_1+E^a_2+E^a_3= 0$ and therefore is parametrized by seven ($5\times 3-8$) quantum numbers. In order to construct such a singlet Hilbert space, it is convenient to go to the Schwinger boson representation of SU(3).

 SU(3) has two inequivalent fundamental representations $3$ and $3^*$ which in the above notation is (1,0) and (0,1) respectively. However, a naive generalization of SU(2) schwinger bosons using a harmonic oscillator triplet and anti-triplet leads  to multiplicity problem\cite{mc,su3rmi} In SU(2), the states (\ref{su2basis}) constructed by taking an arbitrary power of harmonic oscillator creation operators 
%\begin{align}
%|n_1,n_2\rangle=(a^\dagger_1)^{n_1}(a^\dagger_2)^{n_2} |0\rangle.
%\end{align}
covers every SU(2) irrep basis state uniquely and any irrep basis state can be constructed this way. This is no longer true for SU(3). i.e, all positive integer powers of the creation operators $\bar{a}^\dagger,\bar{b}^\dagger$ of $3$ and $3^*$ representation acting on the Schwinger boson vacuum gives the basis of a reducible space \cite{su3rmi} in general. 
%This is because such a construction gives each irrep many times.
 This problem arises due to existence of an SU(3) invariant operator $(\bar{a}^\dagger \cdot \bar{b}^\dagger)$ which acting on any irrep state gives another state which transforms the same way (in contrast, in SU(2), $2$ representation is equivalent to $2^*$ representation and therefore no such invariants exist, for instance $a^\dagger \cdot \tilde{a}^\dagger=0$). One can use the following Sp(2,R) algebra to label \cite{mc} such multiplicities. 
\begin{align}
	k_1&=\frac{1}{2}(\bar{a}^\dagger \cdot \bar{b}^\dagger+\bar{a}\cdot \bar{b})\nonumber\\
	k_2&=-\frac{i}{2}(\bar{a}^\dagger \cdot \bar{b}^\dagger-\bar{a}\cdot \bar{b})\nonumber\\
	k_0&= \frac{1}{2}(\hat{n}+\hat{m}+3)
\end{align}	
where $\hat{n}=a^\dagger \cdot a, \hat{m}=b^\dagger \cdot b$. One can define $k_\pm=k_1 \pm i k_2$ which raises/ lowers $k_0$ eigenvalue as $[k_0,k_\pm]=\pm k_\pm$. The above Sp(2,R) generators commutes with the SU(3) generators. The eigenvalue of $k_0$ distinguishes different multiplicities.  States which satisfies the constraint $k_- |\rangle=0$, where $|\rangle$ is an arbitrary state, are SU(3) irrep states without multiplicites . One can construct irreducible Schwinger bosons \cite{su3rmi} which directly creates SU(3) irrep states without multiplicities by solving the above constraints and is given by : 
\begin{align}
a^\dagger_\alpha=\bar{a}^\dagger_\alpha- \frac{1}{n+m+1}k_+ \bar{b}_\alpha \hspace{1cm}
b^\dagger_\alpha=\bar{b}^\dagger_\alpha-\frac{1}{n+m+1}k_+\bar{a}_\alpha
\end{align}
 
%	    Therefore, in order to create an irreducible representation, one need to use
	     These irreducible SU(3) prepotentials 
%	     triplets $a_\alpha$ and $b_\beta$  which 
	     have the following modified commutation relations (Dirac brackets):   
\begin{align}
\big[a_\alpha~,~a^\dagger_\beta\big]&= \delta_{\alpha\beta}-\tilde{N}~ b^\dagger_\alpha b_\beta \nonumber \\
\big[b_\alpha~,~b^\dagger_\beta\big]&= \delta_{\alpha\beta}-\tilde{N}~ a^\dagger_\alpha a_\beta \nonumber\\
\big[a_\alpha~,~b^\dagger_\beta\big]&= -\tilde{N}~ b^\dagger_\alpha a_\beta
\label{aircom}
\end{align} 
Above, $\tilde{N}=\frac{1}{\hat{N}+2}$ where $\hat{N}=\hat{n}+\hat{m}$ is the number operator for the total number of oscillators. All other possible commutators vanish. Further, they automatically satisfy the following relation: $(a^\dagger\cdot b^\dagger)|\rangle=0$.
%where $|\rangle$ is an arbitrary state. 
\begin{figure}
	\includegraphics[]{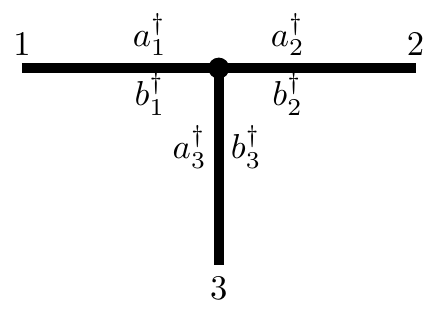}
	\caption{ The three SU(3) irreps are denoted by three lines meeting at a point and  three triplets $a^\dagger$ and three anti-triplets $b^\dagger$ of SU(3) irreducible Schwinger bosons are associated with it.}
	%  		and the triplet indices are suppressed. }
	\label{3vertexsu3}
\end{figure}
SU(3) generators can be written in terms of irreducible SU(3) Schwinger bosons as follows
\begin{align}
E^a= a^\dagger \frac{\lambda^a}{2} a - b^\dagger \frac{\lambda^{*a}}{2} b
\end{align}
where $\lambda^a$ are the eight generators of SU(3) in the 3 representation.

Consider three SU(3) irreps constructed using three pairs of irreducible Schwinger bosons $(a^\dagger_i,b^\dagger_i); i=1,2,3$. These three irreps are denoted pictorially as in figure \ref{3vertexsu3}. A basis of the SU(3) singlet representation can be constructed by taking all the singlet operators constructed from the Schwinger boson creation operators and acting on the Schwinger boson vacuum state as follows:
%A basis of the SU(3) singlet representation constructed from the 3 SU(3) irrep is 
  \begin{align}
|l,p\rangle = \frac{1}{\sqrt{S(l,p)}} |l,p\rangle_u=\begin{cases} \frac{1}{\sqrt{S(l,p)}} \prod\limits_{\stackrel{i=1,2,3}{j=1,2,3}}(a_{i}^\dagger \cdot b_{j} )^{l_{ij}} (\epsilon a^\dagger_1 a^\dagger_2 a^\dagger_3)^{|p|}|0\rangle & p\geq0\nonumber\\
\frac{1}{\sqrt{S(l,p)}} \prod\limits_{\stackrel{i=1,2,3}{j=1,2,3}}(a_{i}^\dagger \cdot b_{j} )^{l_{ij}} (\epsilon b^\dagger_1 b^\dagger_2 b^\dagger_3)^{|p|}|0\rangle & p\leq0 
\end{cases}
\label{su3gib}
\end{align}
where, $l\equiv \{l_{12},l_{21},l_{13},l_{31},l_{23},l_{32}\}; (\epsilon a^\dagger_1 a^\dagger_2 a^\dagger_3) \equiv (\epsilon_{\alpha\beta\gamma} a^\dagger_{1,\alpha} a^\dagger_{2,\beta} a^\dagger_{3,\gamma})$  (clockwise phase convention used); $|l,p\rangle_u$ is the unnormalized basis and the normalization $S(l,p) = {}_u\langle l,p|l,p\rangle_u$ is calculated in appendix \ref{app1}. Orthogonality of the basis can be shown as follows. When $p\geq0$
\begin{align}
\langle l',p'|l,p\rangle=\frac{1}{S(l,p)}\langle0|(\epsilon a_1 a_2 a_3)^{p'} (a_i \cdot b_j)^{l'_{ij}}(a^\dagger_i \cdot b^\dagger_j)^{l_{ij}}(\epsilon a^\dagger_1 a^\dagger_2 a^\dagger_3)^{p}|0\rangle=\delta_{ll'}\delta{pp'}
\end{align}
Last equality is because when $l\neq l'$ or $p\neq p'$, there are extra annihilation or creation operators which hits $|0\rangle$ or $\langle 0|$ to give 0. This is because in the commutation relations (\ref{aircom}), the difference between the total number of annihilation operators and creation operators are preserved in each of the terms. Similarly, orthogonality can be shown for $p<0$ case also. 

\begin{figure*}
	\includegraphics[scale=1.2]{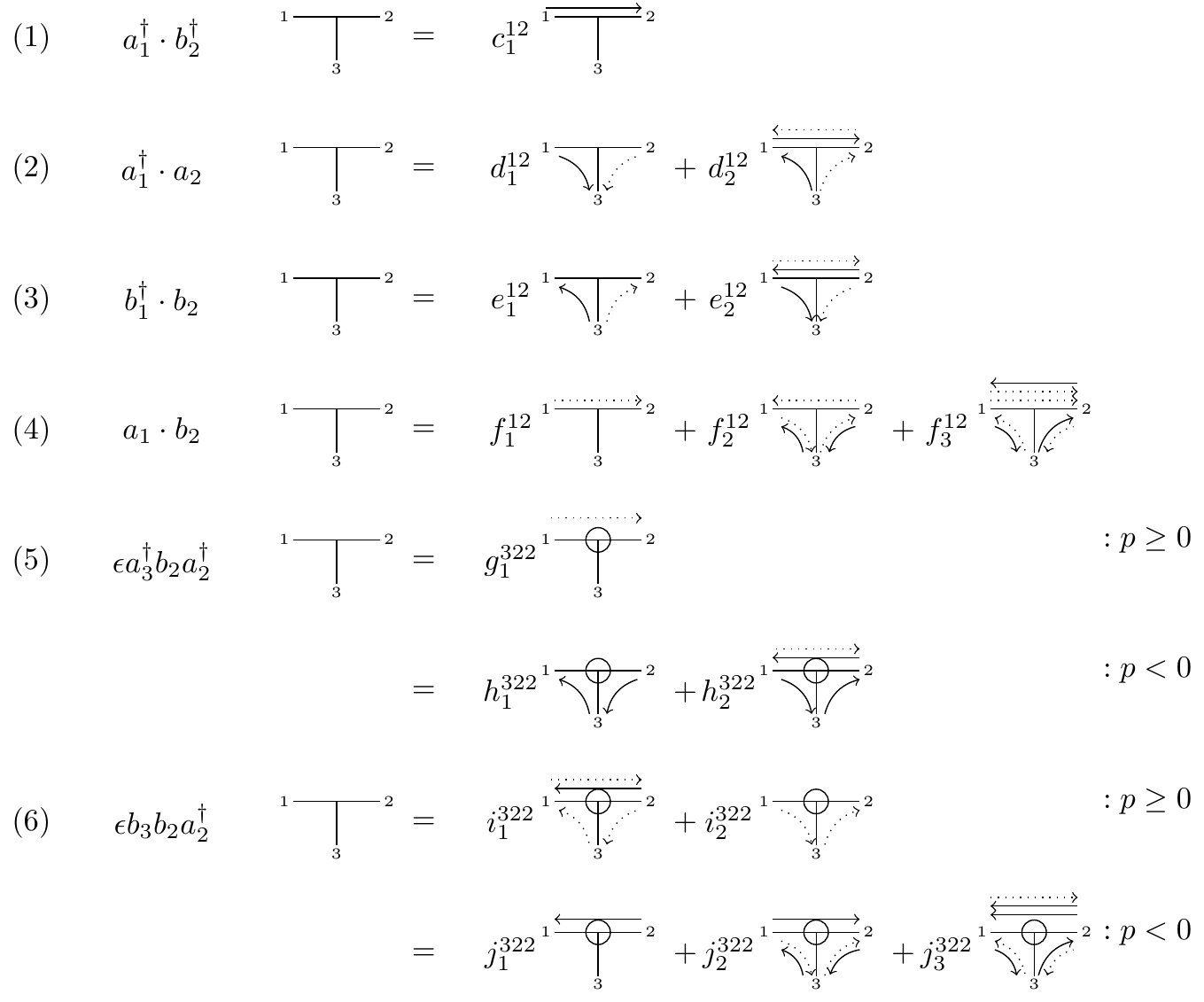}
	\caption{The action of various SU(3) invariant operators on the normalized basis states are shown in the figure. The state $|l,p\rangle$ is denoted by the `$\tt T$' shape. Each solid line from $i$ to $j$ , where $i,j=1,2,3$, denotes increase in the corresponding $l_{ij}$ by 1 and each dashed line denotes decrease by 1. The solid/dashed circle denotes an increase/decrease of p by 1. The coefficients are computed in appendix \ref{app1}. (1),(2),(5) are the basic operator actions from which the action of all other invariant operators can be deduced as described in appendix \ref{app1}.}
	\label{su3opaction}
\end{figure*}

The action of a gauge invariant operator $O$ on a general normalised basis state is 
\begin{align}
O~|l,p\rangle &= \frac{1}{\sqrt{S(l,p)}} O|l,p\rangle_u\nonumber\\
&\equiv \sum\limits_{l',p'} ~\Bigg[\sqrt{\frac{S(l',p')}{S(l,p)}}~\Bigg]~C_{l',p'}~|l',p'\rangle
\end{align}
Therefore, in order to compute the action of various operators on normalized states, one only need the action of those operators on the unnormalized states and ratio of norms. These are calculated in appendix \ref{app1}. 
Complete set of basic invariant operators are given by $(a^\dagger_i \cdot b^\dagger_j), (a^\dagger_i \cdot a_j), (b^\dagger_i\cdot b_j), \Big(\epsilon (a/b^\dagger)_i (a/b^\dagger)_j(a/b^\dagger)_k\Big), \Big(\epsilon (a^\dagger/b)_i (a^\dagger/b)_j(a^\dagger/b)_k\Big)$.
%where is used to denote alternate possible 3 or 3* operators leading to 9 operators. 
Action of various invariant operators on the normalized basis states are  as follows (also pictorially shown in fig. \ref{su3opaction}).
We define $n_1=l_{12}+l_{13},m_1=l_{21}+l_{31},n_2=l_{21}+l_{23},m_2=l_{12}+l_{32}$.
{\footnotesize 
	\begin{enumerate}		
		\item $a^\dagger_1\cdot b^\dagger_2$		
		\begin{align}
(a^\dagger_1\cdot b^\dagger_2)~ |l,p\rangle&=c_1^{12}~|l_{12}+1\rangle\\
c_1^{12}&= \Bigg[\sqrt{\bar{f}_1^{12}(l_{12}+1)}\Bigg] \nonumber
		\end{align}
	\item $(a^\dagger_1\cdot a_2)$
		\begin{align}
&(a^\dagger_1\cdot a_2)~ |l,p\rangle=d_1^{12}~|l_{13}+1,l_{23}-1\rangle
~+~d_2^{12}~\Bigg|{\scriptsize\begin{matrix}l_{12}+1,l_{21}-1\\l_{31}+1,l_{32}-1\\\end{matrix}}\Bigg\rangle
\nonumber
\end{align}
\begin{align}
d_1^{12}&=\Bigg[\sqrt{\frac{{\bar f}_1^{13}(l_{13}+1)\Big|_{l_{12}=l_{21}=0}}{{\bar f}_1^{23}(l_{23})\Big|_{l_{12}=l_{21}=l_{13}=l_{31}=0}}}\Bigg]\frac{(l_{23})(n_2+m_2+|p|-l_{12}+1)}{(n_2+m_2+|p|+1)} \nonumber\\
d_2^{12}&=\Bigg[\sqrt{\frac{{\bar f}_1^{12}(l_{12}+1) ~{\bar f}_1^{31}(l_{31}+1)\Big|_{l_{12}=l_{21}=l_{13}=0}}{{\bar f}_1^{21}(l_{21})\Big|_{l_{12}=0}~{\bar f}_1^{32}(l_{32})\Big|_{l_{12}=l_{21}=l_{13}=l_{31}=l_{23}=0}}}\Bigg]\frac{(-l_{32}l_{21})}{(n_2+m_2+|p|+1)}
\label{aafinal}
\end{align}
\item $(b^\dagger_1\cdot b_2)$
\begin{align}
&(b^\dagger_1\cdot b_2) |l,p\rangle=e_1^{12}~|l_{31}+1,l_{32}-1\rangle
~+~e_2^{12}~\Bigg|{\scriptsize\begin{matrix}l_{21}+1,l_{12}-1\\l_{13}+1,l_{23}-1\end{matrix}}\Bigg\rangle
\nonumber
\end{align}
\begin{align}
e_1^{12}&=\Bigg[\sqrt{\frac{{\bar f}_1^{31}(l_{31}+1)\Big|_{l_{12}=l_{21}=l_{13}=0}}{{\bar f}_1^{32}(l_{32})\Big|_{l_{12}=l_{21}=l_{13}=l_{31}=l_{23}=0}}}\Bigg]\frac{(l_{32})(n_2+m_2+|p|-l_{21}+1)}{(n_2+m_2+|p|+1)}\nonumber\\
e_2^{12}&=\Bigg[\sqrt{\frac{{\bar f}_1^{21}(l_{21}+1)\Big|_{l_{12}=0} ~{\bar f}_1^{13}(l_{13}+1)\Big|_{l_{12}=l_{21}=0}}{{\bar f}_1^{12}(l_{12})~{\bar f}_1^{23}(l_{23})\Big|_{l_{12}=l_{21}=l_{13}=l_{31}=0}}}\Bigg]\frac{(-l_{23}l_{12})}{(n_2+m_2+|p|+1)}
\label{bbfinal}
\end{align}
\item $(a_1\cdot b_2)$
\begin{align}
&(a_1\cdot b_2) |l,p\rangle =  f_1^{12}
|l_{12}-1\rangle + f_2^{12}~
\Bigg|{\scriptsize\begin{matrix}l_{21}-1,\\l_{23}+1,l_{32}-1\\l_{31}+1,l_{13}-1, \end{matrix}}\Bigg\rangle + f_3^{12}~
\Bigg|{\scriptsize\begin{matrix}
l_{21}+1,~l_{12}-2\\l_{13}+1,~l_{31}-1\\l_{32}+1,l_{23}-1
\end{matrix}}\Bigg\rangle
\nonumber
\end{align}
\begin{align}
f_1^{12}&=\Bigg[\sqrt{{\bar{f}_1^{12}(l_{12})}}\Bigg]\nonumber\\
f_2^{12}&=-\Bigg[\sqrt{\frac{{\bar f}_1^{23}(l_{23}+1)\Big|_{l_{12}=l_{21}=l_{13}=l_{31}=0} ~{\bar f}_1^{31}(l_{31}+1)\Big|_{l_{12}=l_{21}=l_{13}=0}}{{\bar f}_1^{21}(l_{21})\Big|_{l_{12}=0}~{\bar f}_1^{32}(l_{32})\Big|_{l_{12}=l_{21}=l_{13}=l_{31}=l_{23}=0}~{\bar f}_1^{13}(l_{13})\Big|_{l_{12}=l_{21}=0}}}\Bigg]\nonumber\\
&(l_{32})(l_{13})l_{21}\bigg(\frac{n_1+m_1+n_2+m_2+2|p|+3 -l_{21}}{(n_1+m_1+|p|+1)(n_2+m_2+|p|+1)}\bigg) \nonumber\\ 
f_3^{12}&=\Bigg[\sqrt{\frac{{\bar f}_1^{32}(l_{32}+1)\Big|_{l_{12}=l_{21}=l_{13}=l_{31}=l_{23}=0} ~{\bar f}_1^{13}(l_{13}+1)\Big|_{l_{12}=l_{21}=0}~{\bar f}_1^{21}(l_{21})\Big|_{l_{12}=0}}{{\bar f}_1^{12}(l_{12}){\bar f}_1^{12}(l_{12}-1)~{\bar f}_1^{31}(l_{31})\Big|_{l_{12}=l_{21}=l_{13}=0}~{\bar f}_1^{23}(l_{23})\Big|_{l_{12}=l_{21}=l_{13}=l_{31}=0}}}\Bigg]\nonumber\\
&~~~\bigg(\frac{(n_1+m_1+|p|+2)(l_{23})(l_{31})}{(n_1+m_1+|p|+1)}\bigg)\frac{l_{12}(l_{12}-1)}{(n_2+m_2+|p|+1)(n_1+m_1+|p|+2)}
\label{finalab}
\end{align}
%above, $n_1=l_{12}+l_{21}+l_{13}+l_{31}+|p|,n_2=l_{12}+l_{21}+l_{23}+l_{32}+|p| $ and
\begin{align} \hspace{-1cm}{\bar{f}_1^{12}(l_{12})}&=\frac{(n_1+m_1+|p|+2)}{(n_1+m_1+|p|+1)}(n_1+m_2-l_{12}+|p|+1)l_{12}-\frac{l_{32}(l_{31}+1)(l_{12})}{n_1+m_1+|p|+1}\bigg(\frac{n_2+m_2+|p|+1-l_{21}}{n_2+m_2+|p|+1}\bigg)\nonumber\\&-\frac{l_{23}(l_{13}+1)(l_{12})}{n_1+m_1+|p|+1}\bigg(\frac{n_1+m_1+|p|+1-l_{21}}{n_2+m_2+|p|+1}\bigg)
\end{align}
\item 
\begin{align}
&(\epsilon a^\dagger_3 b_2 a^\dagger_2)|l,p\rangle=\begin{cases}  g_1^{322}| l_{12}-1,p+1\rangle & p\geq0\\
h_1^{322}\Bigg|{\scriptsize \begin{matrix}p+1\\l_{31}+1,l_{23}+1\end{matrix}}\Bigg\rangle + h_2^{322}
\Bigg|{\scriptsize \begin{matrix}
p+1\\l_{12}-1,l_{21}+1\\l_{13}+1,l_{32}+1\end{matrix}}\Bigg\rangle & p<0
\end{cases} \\
& g_1^{322}=\Bigg[\sqrt{\frac{(|p|+3)^3}{\bar{f}_1^{12}(l_{12})}}\Bigg]l_{12}\nonumber\\
& h_1^{322}=\Bigg[\sqrt{\frac{(\bar{f}_1^{31}(l_{31}+1))\Big|_{l_{12}=l_{21}=l_{13}=0}\bar{f}_1^{23}(l_{23}+1)\Big|_{l_{12}=l_{21}=l_{13}=l_{31}=0}}{(|p|+2)^3}}\Bigg](|p|+l_{12})\nonumber\\
& h_2^{322}=\Bigg[\sqrt{\frac{\bar{f}_1^{21}(l_{21}+1))\Big|_{l_{12}=0}(\bar{f}_1^{13}(l_{13}+1))\Big|_{l_{12}=l_{21}=0}(\bar{f}_1^{32}(l_{32}+1))\Big|_{l_{12}=l_{21}=l_{13}=l_{31}=l_{23}=0}}{(|p|+2)^3\bar{f}_1^{12}(l_{12})\Big|_{l_{12}=l_{21}=l_{13}=l_{31}=0}}}\Bigg]l_{12}\nonumber
\end{align}
\item $(\epsilon b_3b_2a^\dagger_2)$

\begin{align}
% &p>0:\nonumber\\
(\epsilon b_3b_2a^\dagger_2)|l,p\rangle=\begin{cases}i_1^{322}\Bigg|{\scriptsize\begin{matrix}
l_{12}-1,l_{21}+1\\l_{23}-1,l_{31}-1\\p+1\end{matrix}}\Bigg\rangle+i_2^{322}\Bigg|{\scriptsize\begin{matrix}l_{13}-1,l_{32}-1\\p+1\end{matrix}}\Bigg\rangle
&p\geq0\\
j_1^{322}\Bigg|l_{21}+1,p+1\Bigg\rangle+j_2^{322}\Bigg|{\scriptsize\begin{matrix}
	l_{13}-1,l_{31}+1\\l_{32}-1,l_{23}+1\\l_{12}+1,p+1
	\end{matrix}}\Bigg\rangle+j_3^{322} \Bigg|{\scriptsize \begin{matrix}
	l_{12}-1,l_{21}+2\\l_{23}-1,l_{32}+1,p+1\\l_{13}+1,l_{31}-1
	\end{matrix}}\Bigg\rangle &p<0 \end{cases}
\end{align}
where 
\begin{align}
&i_1^{322}= \sqrt{\frac{\bar{f}_1^{21}(l_{21}+1)|_{l_{12}=0}(|p|+3)^3}{\bar{f}_1^{12}(l_{12})\bar{f}_1^{23}(l_{23})|_{l_{12}=l_{21}=l_{13}=l_{31}=0}\bar{f}_1^{31}(l_{31})|_{l_{12}=l_{21}=l_{13}=0}}} \bar{i}_1^{322} ,\nonumber\\
& i_2^{322}= \sqrt{\frac{(|p|+3)^3}{\bar{f}_1^{13}(l_{13})|_{l_{12}=l_{21}=0}\bar{f}_1^{32}(l_{32})|_{l_{12}=l_{21}=l_{13}=l_{31}=l_{23}=0}}} \bar{i}_2^{322} \nonumber\\
& j_1^{322}= \sqrt{\frac{\bar{f}_1^{21}(l_{21}+1)|_{l_{12}=0}}{(|p|+2)^3}} \bar{j}_1^{322} ,\nonumber\\
& j_2^{322}= \sqrt{\frac{\bar{f}_1^{31}(l_{31}+1)|_{l_{12}=l_{21}=l_{13}=0}\bar{f}_1^{23}(l_{23}+1)|_{l_{12}=l_{21}=l_{13}=l_{31}=0}\bar{f}_1^{12}(l_{12}+1)}{\bar{f}_1^{13}(l_{13})|_{l_{12}=l_{21}=0}\bar{f}_1^{32}(l_{32})|_{l_{12}=l_{21}=l_{13}=l_{31}=l_{23}=0}(|p|+2)^3}} \bar{j}_2^{322} \nonumber\\
&j_3^{322}= \sqrt{\frac{\bar{f}_1^{21}(l_{21}+2)|_{l_{12}=0}\bar{f}_1^{21}(l_{21}+1)|_{l_{12}=0}\bar{f}_1^{13}(l_{13}+1)|_{l_{12}=l_{21}=0}\bar{f}_1^{32}(l_{32}+1)|_{l_{12}=l_{21}=l_{13}=l_{31}=l_{23}=0}}{\bar{f}_1^{12}(l_{12})\bar{f}_1^{23}(l_{23})|_{l_{12}=l_{21}=l_{13}=l_{31}=0}\bar{f}_1^{31}(l_{31})|_{l_{12}=l_{21}=l_{13}=0}(|p|+3)^3}} \bar{j}_3^{322}
\end{align}
where, $\bar{i}_1^{322},\bar{i}_2^{322},\bar{j}_1^{322},\bar{j}_2^{322},\bar{j}_3^{322}$ are given in (\ref{ijbar}).
\item ($\epsilon a^\dagger_1 a^\dagger_2 a^\dagger_3$)
\begin{align}
(\epsilon a^\dagger_1 a^\dagger_2 a^\dagger_3) |l,p\rangle =\begin{cases} n_1^{123} |p+1\rangle & p\geq0 \\
m_1^{123} \Bigg|{\scriptsize \begin{matrix} l_{12}+1, l_{23}+1\\l_{31}+1,p+1 \end{matrix}}\Bigg\rangle + m_2^{123} \Bigg|{\scriptsize \begin{matrix}l_{21}+1,l_{32}+1\\l_{13}+1,p+1\end{matrix}}\Bigg\rangle & p<0 \end{cases}
\end{align}
where 
\begin{align}
& n_1^{123}= \sqrt{(|p|+3)^3} \nonumber\\
& m_1^{123}= \sqrt{\frac{\bar{f}_1^{12}(l_{12}+1)\bar{f}_1^{23}(l_{23}+1)|_{l_{12}=l_{21}=l_{13}=l_{31}=0}\bar{f}_1^{31}(l_{31}+1)|_{l_{12}=l_{21}=l_{13}=0}}{(|p|+2)^3}}\nonumber\\
&m_2^{123}=\sqrt{\frac{\bar{f}_1^{21}(l_{21}+1)|_{l_{12}=0}\bar{f}_1^{32}(l_{32}+1)|_{l_{12}=l_{21}=l_{13}=l_{31}=l_{23}=0}\bar{f}_1^{13}(l_{13}+1)|_{l_{12}=l_{21}=0}}{(|p|+2)^3}}
\end{align}
\end{enumerate}}
In the above expressions, 
\begin{align}
&\bar{f}_1^{21}(l_{21})|_{l_{12}=0}= \frac{l_{21}+l_{13}+l_{31}+|p|+2}{l_{21}+l_{13}+l_{31}+|p|+1}(l_{21}+l_{31}+l_{23}+|p|+1)l_{21}-\frac{l_{21}l_{23}(l_{13}+1)}{l_{21}+l_{13}+l_{31}+1}\nonumber\\
&-\frac{l_{21}l_{32}(l_{31}+1)}{l_{21}+l_{23}+l_{32}+1} \nonumber\\
& \bar{f}_1^{13}(l_{13})|_{l_{12}=l_{21}=0}=\Big(\frac{l_{13}+l_{31}+l_{23}+l_{32}+|p|+2}{l_{13}+l_{31}+l_{23}+l_{32}+|p|+1}\Big)(l_{13}+l_{23}+|p|+1)l_{13}\nonumber \\
& \bar{f}_1^{31}(l_{31})|_{l_{12}=l_{21}=l_{13}=0}= \Big(\frac{l_{31}+l_{23}+l_{32}+|p|+2}{l_{31}+l_{23}+l_{32}+|p|+1}\Big)(l_{31}+l_{32}+|p|+1)l_{31}\nonumber\\
& \bar{f}_1^{23}(l_{23})|_{l_{12}=l_{21}=l_{13}=l_{31}=0}=\bigg(\frac{l_{23}+l_{32}+|p|+2}{l_{23}+l_{32}+|p|+1}\bigg)(l_{23}+|p|+1)l_{23} \nonumber\\
&\bar{f}_1^{32}(l_{32})|_{l_{12}=l_{21}=l_{13}=l_{31}=l_{23}=0}= \Big(\frac{l_{32}+|p|+2}{l_{32}+|p|+1}\Big)(l_{32}+|p|+1)l_{32}
\end{align}

 In the above expressions all $l_{ij}$ are numbers labelling the L.H.S ket vector. The ket vectors in R.H.S are the same as the L.H.S, only the changed labels are mentioned in the notation.
 $\bar{f}_1^{21},\bar{f}_1^{13},\bar{f}_1^{31},\bar{f}_1^{23},\bar{f}_1^{32}$ can be obtained from $\bar{f}_1^{12}$ by the two symmetry operations described below.
Since $a$ and $b$ are in the same footing in the basis states and three SU(3) irreps are also in the same footing, one can make the following operations on the equation giving the action of various operators and arrive at another valid equation: 
%there are the following symmetries in the action of the operators
\begin{enumerate}
	\item 
	simultaneously perform the following operations:
	\begin{itemize}
	\item $a\leftrightarrow b$ 
	\item $l_{ij} \leftrightarrow l_{ji}$ 
	\item  $p>0$ case and $p<0$ case interchanged.
	\item where ever p increases by some value in the state, replace it by p decreasing by the same value and vice versa.
\end{itemize}
%	   in the  p increasing by 1 and p decreasing by 1 interchanged.
	\item cyclic permutation of $(1,2,3)$
\end{enumerate}
%action of the operator constructed by the interchange of $b$ and $a$ in a gauge invariant operator is given by making the transformating $l_{\alpha \beta} \rightarrow l_{\beta \alpha}$. 
For example, in Figure \ref{su3opaction},  (3) is obtained from (2) by $a\leftrightarrow b, l_{ij} \leftrightarrow l_{ji}$.
%just flipping the direction of all the arrows and making $l_{\alpha \beta}\rightarrow l_{\beta \alpha}$ in the coefficients.
% These symmetries can be used to deduce the action of all other invariant operators from fig. \ref{su3opaction}. 

Action of all other invariant operators can be easily deduced either by using the above symmetries or by reducing those operators in terms of the basic operators given in fig.\ref{su3opaction}. We will illustrate this for the operators $a_1\cdot b_2, \epsilon b_3b_2a^\dagger_2, \epsilon a_3a_2b^\dagger_2$ and $ \epsilon a_1a_2a_3$ in appendix \ref{app1}.
% $(\epsilon_{rst}b_{3,r}b_{2,s}a^\dagger_{2,t})$. 
% $(\epsilon_{rst}b_{3,r}b_{2,s}a^\dagger_{2,t}) |l,p\rangle_u= \sum\limits_{l',p'} C_{l'p'} |l',p'\rangle_u$. We can calculate $C_{l'p'}$ as follows: 
%  $\Big[b_3\cdot b^\dagger_1,\epsilon a^\dagger_3b_2a^\dagger_2\Big]|l,p\rangle=\sum\limits_{\bar{l},\bar{p}} D_{\bar{l},\bar{p}} |\bar{l},\bar{p}\rangle_u=-\tilde{N}_3(a^\dagger_3\cdot b^\dagger_1)(\epsilon b_3b_2a^\dagger_2)|l,p\rangle_u=\frac{-1}{n_3+3}\sum\limits_{l',p'} C_{l',p'}|l'_{31}+1,l',p'\rangle_u$. We know the action of $b_3\cdot b^\dagger_1$ and $\epsilon a^\dagger_3b_2a^\dagger_2$ and therefore $D_{\bar{l}\bar{p}}$ from figure (\ref{su3opaction}). Therefore, $C_{l,p}=-(n_3+2)D_{l_{32}+1,l,p}$.
% \begin{align}
% \end{align}

%\subsection{Fermionic representation}
%\label{su3ferm}
\section{Concluding remarks}
\label{con}
We have studied the problem of constructing a singlet Hilbert space from the direct product of three arbitrary SU(3) irreps and calculating the action of invariant operators on this space. Singlet spaces are of importance when the Hilbert space has to satisfy gauge invariance constraints. In order to study the dynamics of such systems one requires the action of Hamiltonian on such a Hilbert space. Hamiltonian can be written in terms of the invariant operators. Therefore, one needs to evaluate the action of such basic invariant operators on the basis states. For SU(3), we have evaluated the action of these operators. In particular, the construction described in this paper is useful in defining the physical Hilbert space of lattice gauge theories and studying its dynamics.

One can consider product of Hilbert spaces defined here. For example, doing a product of two such Hilbert space with one common casimir will generate a diagram $>-<$ where the intermediate casimir is identified to be the same. This kind of product can be done in three different ways. All the three Hilbert spaces are unitary transforms of each other. Indeed the transformation coefficients are the 6-j symbols in the case of SU(2). These 6-j symbols depend only on the invariant casimir quantum numbers. We comment that the most primitive Hilbert space which only depend on casimirs is the singlet Hilbert space considered here. 
%The states we have defined in this paper are the simplest 'primitive' states depending only on the invariant 'casimir' quantum numbers. One can calculate higher wigner symbols , both for SU(2) as well as SU(3), as the transformations between the direct product of such primitive states constructed in different ways.

 One can generalize these results to SU(N) again by using SU(N) irreducible \cite{sunb} Schwinger bosons. Further, another generalization which is of use, is to replace one or more of Schwinger boson pairs by fermionic oscillator pairs. This corresponds to having Gauss law in the presence of charged currents such as quarks. It will be considered in a future publication.

%\paragraph{Fourth-level heading is run in.}%

\appendix
%\section{SU(2): Calculation of action of various invariant operators and norm}
%\subsection{Schwinger boson representation}
%\subsection{fermionic representation}
\section{SU(3):Calculation of action of various invariant operators on the unnormalized basis}
\label{app1}
The action of a gauge invariant operator $O$ on the $|l,p\rangle_u$ can be computed by using the commutation relations (\ref{aircom}) to shift $O$ across $(a^\dagger_i \cdot b^\dagger_j)$ and $(\epsilon_{\alpha\beta\gamma}a^\dagger_{i,\alpha} a^\dagger_{j,\beta} a^\dagger_{k,\gamma})$ or $(\epsilon_{\alpha\beta\gamma}b^\dagger_{i,\alpha} b^\dagger_{j,\beta} b^\dagger_{k,\gamma})$ until it hits $|0\rangle$.	In the following, whenever there is no confusion, only the $l_{ij},p$ values which are different from the $l,p$ are written. Also, it is convenient to introduce the following notation $(i\cdot j)\equiv a^\dagger_i\cdot b^\dagger_j, N_i\equiv n_i+m_i+|p|$.
% This procedure is illustrated for $O=  a^\dagger_1\cdot a_2$ and action of other operators are then stated. 
\begin{enumerate}
	\item $(a^\dagger_1\cdot a_2)$
	\begin{align}
	(a^\dagger_1\cdot a_2) | l,p\rangle_u = (a^\dagger_1\cdot a_2) (a^\dagger_1\cdot b^\dagger_2) |l_{12}-1\rangle_u
	\end{align}

%	$\bar{l}_{\alpha\beta}$ denotes all the $l$ quantum numbers not mentioned explicitly.
	 Using (\ref{aircom}), 
	\begin{align}
	(a^\dagger_1\cdot a_2) (a^\dagger_1\cdot b^\dagger_2\Big)
	%	&=\Big(a^\dagger[1]\cdot b^\dagger[2]\Big)\Big(a^\dagger[1]\cdot a[2]\Big)-\tilde{N}[2]\Big(a^\dagger[1]\cdot b^\dagger[2]\Big)\Big(a^\dagger[1]\cdot a[2]\Big)\nonumber\\
	&=(1-\tilde{N}_2)(a^\dagger_1\cdot b^\dagger_2)(a^\dagger_1\cdot a_2)
	\end{align}
	Repeating this $l_{12}$ times we get, 
	\begin{align}
	(a^\dagger_1\cdot a_2)|
	l,p\rangle_u&=(1\cdot2)^{l_{12}}\left(\frac{\hat{N}_2+2}{\hat{N}_2+l_{12}+2}\right)(a^\dagger_1\cdot a_2)|l_{12}=0\rangle_u
	\label{I1it}
	\end{align}
	Using the relation 
%	\begin{align}
$	(a^\dagger_1\cdot a_2) (a^\dagger_2\cdot b^\dagger_1)=(a^\dagger_2\cdot b^\dagger_1)(a^\dagger_1\cdot a_2) -\tilde{N}_2(a^\dagger_1\cdot b^\dagger_2)( b^\dagger_1\cdot b_2)$
%	\end{align}
	repeatedly, we get 
	\begin{align}
		(a^\dagger_1\cdot a_2) |l_{12}=0\rangle_u &= (a^\dagger _2 \cdot b^\dagger_1)^{l_{21}} (a^\dagger_1\cdot a_2)|l_{12}=0,l_{21}=0\rangle_u\nonumber\\ 
	&+ \sum\limits_{r=0}^{l_{21}-1}(a^\dagger_2 \cdot b^\dagger_1)^r \bigg\{-\tilde{N}_2(a^\dagger_1\cdot b^\dagger_2)( b^\dagger_1\cdot b_2) \bigg\}|l_{12}=0,l_{21}-1-r\rangle_u
	\label{Iit2pre}
	\end{align}
	Using, $(b^\dagger_1\cdot b_2)(2\cdot 1)= (1-\tilde{N}_2)(2\cdot1)(b^\dagger_1\cdot b_2)$, we get 
	\begin{align}
	( b^\dagger_1\cdot b_2) |l_{12}=0,l_{21}-1-r\rangle_u=(2\cdot 1)^{l_{21}-1-r} \bigg(\frac{\hat{N}_2+2}{\hat{N}_2+l_{21}+1-r}\bigg)(b^\dagger_1\cdot b_2)|l_{12}=0,l_{21}=0\rangle_u
	\end{align}
	We now calculate $(a^\dagger_1\cdot a_2)~|l_{12}=0,l_{21}=0\rangle_u $. Using 
	$(a^\dagger_1\cdot a_2)(a^\dagger_3\cdot b^\dagger_2)=(a^\dagger_3\cdot b^\dagger_2)(a^\dagger_1\cdot a_2) -\tilde{N}_2 (a^\dagger_1\cdot b^\dagger_2)(a^\dagger_3\cdot a_2)$, we get:
	\begin{align}
	&(a^\dagger_1\cdot a_2)|l_{12}=0,l_{21}=0\rangle_u=(3 \cdot 2)^{l_{32}}(a^\dagger_1\cdot a_2)|l_{12}=0,l_{21}=0,l_{32}=0\rangle_u+ \nonumber \\
	&\sum\limits_{r=0}^{l_{32}-1}(3 \cdot 2)^r \bigg(\frac{-1}{\hat{N}_2+2}\bigg)(1\cdot 2)\bigg\{\Big(\frac{\hat{N}_2+2}{\hat{N}_2+1}\Big)(3\cdot 2)\bigg\}^{l_{32}-1-r}(a^\dagger_3\cdot a_2)|l_{12}=0,l_{21}=0,l_{32}=0\rangle_u
	\label{I2it}
	\end{align}
	Since, $(a^\dagger_3 \cdot a_2)|l_{12}=0,l_{21}=0,l_{32}=0\rangle_u=0$ and $a^\dagger_1\cdot a_2)(a^\dagger_2\cdot b^\dagger_3)=(1 \cdot 3)+(2\cdot 3) (a^\dagger_1 \cdot a_2)-\tilde{N}_2(1 \cdot 2)(b^\dagger_3\cdot b_2)$, we get 
	\begin{align}
(a^\dagger_1 \cdot a_2)|l_{12}=0,l_{21}=0,l_{32}=0\rangle_u= l_{23} (1\cdot 3) |l_{12}=0,l_{21}=0,l_{32}=0,l_{23}-1\rangle_u
	\label{Iit41}
	\end{align} 
	Putting this back into (\ref{I2it}) gives: 
	\begin{align}
(a^\dagger_1\cdot a_2)&|l_{12}=0,l_{21}=0\rangle_u=l_{23}|l_{12}=0,l_{21}=0,l_{23}-1,l_{13}+1 \rangle_u
	\end{align}
	We now calculate $(b^\dagger_1\cdot b_2)~|l_{12}=0,l_{21}=0\rangle_u $:
	Since, 
	\begin{align}
(b^\dagger_1\cdot b_2) (a^\dagger_2\cdot b^\dagger_3)=(2\cdot 3) (b^\dagger_1\cdot b_2)-\tilde{N}_2 (2\cdot 1)(b^\dagger_3\cdot b_2),
	\end{align}
	we have,
	\begin{align}
	(b^\dagger_1\cdot b_2)|l_{12}=0,l_{21}=0\rangle_u&=(2\cdot 3)^{l_{23}}(b^\dagger_1\cdot b_2)|l_{12}=0,l_{21}=0,l_{23}=0\rangle_u + \nonumber \\ 
	&\sum\limits_{r=0}^{l_{23}-1} (2\cdot 3)^r \bigg\{-\tilde{N}_2 (2\cdot 1) (b^\dagger_3\cdot b_2)\bigg\}|l_{12}=0,l_{21}=0,l_{23}-1\rangle_u
	\end{align}
	Since, $(b^\dagger_3\cdot b_2)(2\cdot 3)= \Big(\frac{N_2+1}{N_2+2}) (2\cdot 3) (b^\dagger_3\cdot b_2)$ and 
	\begin{align}
	&(b^\dagger_3\cdot b_2) |l_{12}=0,l_{21}=0,l_{23}=0\rangle_u =(b^\dagger_3\cdot b_2)(a^\dagger_3\cdot b^\dagger_2)|l_{12}=0,l_{21}=0,l_{23}=0,l_{32}-1\rangle_u\nonumber\\
	&=(3\cdot 2)^{l_{32}}(b^\dagger_3\cdot b_2)|l_{12}=0,l_{21}=0,l_{23}=0,l_{32}=0\rangle_u\nonumber\\
	&+\sum_{r=0}^{l_{32}-1}(3\cdot 2)^r \bigg\{-\tilde{N}_2(2\cdot 3)(a^\dagger_3\cdot a_2)\bigg\} |l_{12}=0,l_{21}=0,l_{23}=0,l_{32}-1-r\rangle_u=0,
	\end{align} 
	 we have 
	\begin{align}
	(b^\dagger_1\cdot b_2)|l_{12}=0,l_{21}=0\rangle_u&=(2\cdot 3)^{l_{23}}(b^\dagger_1\cdot b_2) |l_{12}=0,l_{21}=0,l_{23}=0\rangle_u
	\end{align}
	Now, $(b^\dagger_1\cdot b_2) |l_{12}=0,l_{21}=0,l_{23}=0\rangle_u=l_{32}(3\cdot 1)|l_{12}=0,l_{21}=0,l_{23}=0,l_{32}-1\rangle_u$. Therefore, 
	\begin{align}
(b^\dagger_1\cdot b_2)|l_{12}=0,l_{21}=0\rangle_u=l_{32}|l_{12}=0,l_{21}=0,l_{32}-1,l_{31}+1\rangle_u
	\label{IIit42}
	\end{align}
	Putting (\ref{Iit41}) and (\ref{IIit42}) back in (\ref{Iit2pre}), we get 
{\footnotesize
		\begin{align}
	&(a^\dagger_1\cdot a_2) |l,p\rangle_u=(1.2)^{l_{12}}(2\cdot 1)^{l_{21}}\bigg(\frac{\hat{N}_2+l_{21}+2}{\hat{N}_2+l_{21}+l_{12}+2}\bigg)l_{23}|l_{12}=0,l_{21}=0,l_{31}+1,l_{23}-1\rangle_u\nonumber\\
	&+(1\cdot 2)^{l_{12}+1}(2\cdot 1)^{l_{21}-1}\bigg(\frac{(\hat{N}_2+l_{21}+2)(\hat{N}_2+2)}{\hat{N}_2+l_{12}+l_{21}+2}\bigg)\sum\limits_{r=0}^{l_{21}-1}\bigg(\frac{-l_{32}}{(\hat{N}_2+l_{21}+2-r)(\hat{N}_2+l_{21}+1-r)}\bigg)\nonumber\\ 
	&\hspace{2.7cm}\Big|l_{12}=0,l_{21}=0,l_{32}-1,l_{31}+1\Big\rangle_u
	\end{align}}
	Summing over $r$ and simplifying gives:
%	Repeating the same procedure to shift $a^\dagger_1\cdot a_2$ to the right until it hits $|0\rangle_u$, we get :
%	\begin{widetext} 
		\begin{align}
		&(a^\dagger_1\cdot a_2)~ |l,p\rangle_u={\bar d}_1^{12}~\Bigg|\begin{matrix}l_{13}+1,l_{23}-1\end{matrix}\Bigg\rangle_u
		~+~{\bar d}_2^{12}~\Bigg|\begin{matrix}l_{12}+1,l_{21}-1\\l_{31}+1,l_{32}-1\end{matrix}\Bigg\rangle_u \nonumber
		\end{align}
		\begin{align}
		{\bar d}_1^{12}=\frac{(l_{23})(n_2+m_2-l_{12}+|p|+1)}{(n_2+m_2+|p|+1)} \hspace{1cm}
		{\bar d}_2^{12}=\frac{(-l_{32}l_{21})}{(n_2+m_2+|p|+1)}
		\label{aafinal}
		\end{align}
		\item $(b^\dagger_1\cdot b_2)$:
		
		After a similar calculation,
		\begin{align}
		&(b^\dagger_1\cdot b_2) |l,p\rangle_u={\bar e}_1^{12}~|\begin{matrix}l_{31}+1,l_{32}-1\end{matrix}\rangle_u
		~+~{\bar e}_2^{12}~\Bigg|\begin{matrix}l_{21}+1,l_{12}-1,\\l_{13}+1,l_{23}-1\end{matrix}\Bigg\rangle_u\nonumber
		\end{align}
		\begin{align}
		\bar{e}_1^{12}=\frac{(l_{32})(n_2+m_2-l_{21}+|p|+1)}{(n_2+m_2+|p|+1)}\hspace{1cm}
		\bar{e}_2^{12}=\frac{(-l_{23}l_{12})}{(n_2+m_2+|p|+1)}
			\label{bbfinal}
		\end{align}
		\item $(a_1\cdot b_2)$
		
		Using the following relation: 
		\begin{align}
		&(a_{1,\alpha} b_{2,\alpha})~(a_{1,\beta}^\dagger b_{2,\beta}^\dagger)= a_{1,\alpha}a^\dagger_{1,\beta}b^\dagger_{2,\beta}b_{2,\alpha}+a_1\cdot a^\dagger_1-\tilde{N}_2(a_1\cdot a^\dagger_2)(a^\dagger_1\cdot a_2)\nonumber\\
		&=(a^\dagger_1\cdot b^\dagger_2)(a_1\cdot b_2)+b^\dagger_2\cdot b_2-\tilde{N}_1b^\dagger_{1,\alpha}b_{1,\beta}b^\dagger_{2,\beta}b_{2,\alpha}+a_1\cdot a^\dagger_1-\tilde{N}_2(a_1\cdot a^\dagger_2)(a^\dagger_1\cdot a_2)\nonumber\\
		&=(a^\dagger_1\cdot b^\dagger_2)(a_1\cdot b_2)+b^\dagger_2\cdot b_2-\tilde{N}_1\bigg(b_{1,\beta}b^\dagger_{1,\alpha}b^\dagger_{2,\beta}b_{2,\alpha}-b^\dagger_2\cdot b_2+\tilde{N}_1a^\dagger_{1,\beta}a_{1,\alpha}b^\dagger_{2,\beta}b_{2,\alpha}\bigg)\nonumber\\
		&\hspace{3cm}+a_1\cdot a^\dagger_1-\tilde{N}_2( a^\dagger_2\cdot a_1)(a^\dagger_1\cdot a_2)\nonumber
		\end{align} repeatedly, we get
		{\footnotesize
		\begin{align}
		&(a_1\cdot b_2)~|l,p\rangle_u=\Big[(1-\tilde{N}_1^2)(1\cdot 2)\Big]^{l_{12}} (a_1\cdot b_2) |l_{12}=0\rangle+\sum\limits_{r=0}^{1_{12}-1}\Big[(1-\tilde{N}_1^2)(1\cdot 2)\Big]^r\nonumber\\&\bigg\{3+\hat{N}_{2b}+\hat{N}_{1a}+\tilde{N}_1(\hat{N}_{2b}-\hat{N}_{1b})-\tilde{N}_1(b^\dagger_2\cdot b_1)(b^\dagger_1\cdot b_2)-\tilde{N}_2( a^\dagger_2\cdot a_1)(a^\dagger_1\cdot a_2)\bigg\}|l_{12}-1-r\rangle_u
		\label{1it}
		\end{align}}
		Using, 
		\begin{align}
		&(a_1\cdot b_2)(a^\dagger_2\cdot b^\dagger_1)=a_{1,\alpha}a^\dagger_{2,\beta}b_{2,\alpha}b^\dagger_{1,\beta}-\tilde{N}_2(a^\dagger_2\cdot a_1)(b^\dagger_1\cdot b_2)\nonumber\\
		&=a^\dagger_{2,\beta}b_{2,\alpha}b^\dagger_{1,\beta}a_{1,\alpha}-\tilde{N}_1b^\dagger_{1,\alpha}a^\dagger_{2,\beta}b_{2,\alpha}a_{1,\beta}-\tilde{N}_2(a^\dagger_2\cdot a_1)(b^\dagger_1\cdot b_2)\nonumber\\
		&=(2\cdot 1)(a_1\cdot b_2)-\tilde{N}_1\Big[(a^\dagger_2,\cdot a_1)(b^\dagger_1 \cdot b_2)+\tilde{N}_1(2\cdot 1)(a_1\cdot b_2)\Big]-\tilde{N}_2( a^\dagger_2\cdot a_1)(b^\dagger_1\cdot b_2)\nonumber
		\end{align}
		repeatedly we get,
		\begin{align}
		&(a_1\cdot b_2) |l_{12}=0,l_{21}\rangle_u=\Big[\Big(1-\tilde{N}_1^2\Big)(2\cdot 1)\Big]^{l_{21}}(a_1\cdot b_2) |l_{12}=0,l_{21}=0\rangle_u \nonumber\\
		&+\sum_{r=0}^{l_{21}-1}\Big[\Big(1-\tilde{N}_1^2\Big)(2\cdot 1)\Big]^r\bigg\{-(\tilde{N}_1+\tilde{N}_2)(a^\dagger_2\cdot a_1)(b^\dagger_1\cdot b_2)\bigg\}|l_{12}=0,l_{21}-1-r\rangle_u
		\label{I2pre1}
		\end{align}
		Since,  $(a_1\cdot b_2)|l_{12}=0,l_{21}=0\rangle=0$, putting back (\ref{I2pre1}) into (\ref{1it}) gives
		{\footnotesize
		\begin{align}
		&(a_1\cdot b_2)|l,p\rangle_u=\nonumber\\
		&-\Big[(1-\tilde{N}_1^2)(1\cdot 2)\Big]^{l_{12}} \sum_{r=0}^{l_{21}-1}\Big[\Big(1-\tilde{N}_1^2\Big)(2\cdot 1)\Big]^r(\tilde{N}_1+\tilde{N}_2)(a^\dagger_2\cdot a_1)(b^\dagger_1\cdot b_2)|l_{12}=0,l_{21}-1-r\rangle_u\nonumber\\&+\sum\limits_{r=0}^{1_{12}-1}\Big[(1-\tilde{N}_1^2)(1\cdot 2)\Big]^r\nonumber\\&\bigg\{3+\hat{N}_{2b}+\hat{N}_{1a}+\tilde{N}_1(\hat{N}_{2b}-\hat{N}_{1b})-\tilde{N}_1(b^\dagger_2\cdot b_1)(b^\dagger_1\cdot b_2)-\tilde{N}_2( a^\dagger_2\cdot a_1)(a^\dagger_1\cdot a_2)\bigg\}|l_{12}-1-r\rangle_u\nonumber
		\end{align}}
		Using (\ref{aafinal}) and (\ref{bbfinal}) and summing over r, we get 
		\begin{align}
		&(a_1\cdot b_2) |l,p\rangle_u =  \bar{f}_1^{12}
	|\begin{matrix}l_{12}-1\end{matrix}\rangle_u + \bar{f}_2^{12}~
		\Bigg|{\scriptsize\begin{matrix}l_{31}+1,l_{21}-1,\\l_{23}+1,l_{32}-1\\l_{13}-1, ~l_{12},~p\end{matrix}}\Bigg\rangle_u + \bar{f}_3^{12}~
		\Bigg|{\scriptsize \begin{matrix}
		l_{21}+1,~l_{12}-2\\l_{13}+1,~l_{31}-1\\l_{23}-1,l_{32}+1,p
		\end{matrix}}\Bigg\rangle_u
		\label{finalab}
		\end{align}
		{\footnotesize
		\begin{align}
	  &\bar{f}_1^{12}(l_{12})=\Bigg[\frac{(n_1+m_1+|p|+2)}{(n_1+m_1+|p|+1)}(n_1+m_2-l_{12}+p+1)l_{12}	\nonumber\\
	 &-\frac{l_{32}(l_{31}+1)(l_{12})}{n_1+m_1+|p|+1}\bigg(\frac{n_2+m_2+|p|+1-l_{21}}{n_2+m_2+|p|+1}\bigg)
	-\frac{l_{23}(l_{13}+1)(l_{12})}{n_1+m_1+|p|+1}\bigg(\frac{n_1+m_1+|p|+1-l_{21}}{n_2+m_2+|p|+1}\bigg)\Bigg]\nonumber\\
&		\bar{f}_2^{12}(l_{12})=-(l_{32})(l_{13})l_{21}\bigg(\frac{n_1+m_1+|p|+n_2+m_2+|p|+3 -l_{21}}{(n_1+m_1+|p|+1)(n_2+m_2+|p|+1)}\bigg) \nonumber\\
		& \bar{f}_3^{12}(l_{12})=\bigg(\frac{(l_{23})(l_{31})}{(n_1+m_1+|p|+1)}\bigg)\frac{l_{12}(l_{12}-1)}{(n_2+m_2+|p|+1)}
		\end{align}}
%	\end{widetext}
	\item $\epsilon a^\dagger_3 b_2 a^\dagger_2$

\begin{itemize} \item When $p\geq0$
	
	Since $\Big[\epsilon a^\dagger_3 b_2 a^\dagger_2,\epsilon a^\dagger_1 a^\dagger_2a^\dagger_3\Big]=0$, we have
	{\footnotesize 
		\begin{align}
		(\epsilon a^\dagger_3 b_2 a^\dagger_2)|l,p\rangle_u&=(\epsilon a^\dagger_1 a^\dagger_2a^\dagger_3)^p(2\cdot 1)^{l_{21}}(1\cdot 3)^{l_{13}}(3\cdot 1)^{l_{31}}(2\cdot 3)^{l_{23}}(3\cdot 2)^{l_{32}} \nonumber\\
		&~~~~~(\epsilon a^\dagger_3 b_2 a^\dagger_2)|p=0,l_{12},l_{\bar{i}\bar{j}}=0\rangle_u
		\end{align}}
	where $l_{\bar{i}\bar{j}}$ denote $l_{21},l_{13},l_{31},l_{23},l_{32}$. Now, we use $(\epsilon a^\dagger_3 b_2 a^\dagger_2)( a^\dagger_1\cdot b^\dagger_2)=(\epsilon a^\dagger_3 a^\dagger_1a^\dagger_2)+( a^\dagger_1\cdot b^\dagger_2)(\epsilon a^\dagger_3 b_2 a^\dagger_2)$ to get 
	\begin{align}
	(\epsilon a^\dagger_3 b_2 a^\dagger_2 )|l_{12},p=0,l_{\bar{i}\bar{j}}=0\rangle_u=l_{12}|l_{12}-1,p=1\rangle_u
	\end{align}
	Therefore, 
	\begin{align}
	(\epsilon a^\dagger_3 b_2 a^\dagger_2)|l,p\rangle_u=l_{12} | l_{12}-1,p+1\rangle_u : p\geq0
	\label{adbadp+}
	\end{align}
	\item When $p<0$
	
	Since $(\epsilon a^\dagger_3 b_2 a^\dagger_2)(\epsilon b^\dagger_1 b^\dagger_2 b^\dagger_3)$, we have 
	{\footnotesize 
		\begin{align}
	&(\epsilon a^\dagger_3 b_2 a^\dagger_2)|l,p\rangle_u=\sum\limits_{r=0}^{|p|-1} (\epsilon b^\dagger_1 b^\dagger_2 b^\dagger_3)^r(3\cdot 1) (2 \cdot 3) |p+r+1\rangle_u+(\epsilon b^\dagger_1 b^\dagger_2 b^\dagger_3)^{|p|} (a^\dagger_3b_2a^\dagger_2)|p=0,l\rangle_u\nonumber\\
	&= |p|(3\cdot 1) (2 \cdot 3) |p+1\rangle_u+ l_{12}(\epsilon b^\dagger_1 b^\dagger_2 b^\dagger_3)^{|p|} (\epsilon a^\dagger_1 a^\dagger_2 a^\dagger_3)|p=0,l_{12}-1\rangle_u
	\end{align}}
	Since $(\epsilon a^\dagger_1a^\dagger_2a^\dagger_3)(\epsilon b^\dagger_1b^\dagger_2b^\dagger_3) =(1\cdot 2)(2\cdot 3)(3\cdot 1)+(2\cdot 1)(1\cdot 3)(3\cdot 2)$, we have
	\begin{align}
	(\epsilon a^\dagger_3 b_2 a^\dagger_2)|l,p\rangle_u=(|p|+l_{12})\Bigg|{\scriptsize\begin{matrix}p+1\\l_{31}+1,l_{23}+1\end{matrix}}\Bigg\rangle_u + l_{12}
	\Bigg|{\scriptsize\begin{matrix}
		p+1\\l_{12}-1,l_{21}+1\\l_{13}+1,l_{32}+1\end{matrix}}\Bigg\rangle_u
	\label{ad3b2ad2p-}
	\end{align}
	Therefore,
	\begin{align}
	(\epsilon a^\dagger_3b_2a^\dagger_2)|l,p\rangle=\begin{cases}\bar{g}_1^{322}|l_{12}-1,p+1\rangle_u & p\geq1\\
	\bar{h}_1^{322} \bigg|{\scriptsize\begin{matrix}
		l_{31}+1,l_{23}+1 \\p+1
		\end{matrix}}\bigg\rangle_u + \bar{h}_2^{322} \bigg|{\scriptsize\begin{matrix}
		l_{12}-1,l_{21}+1\\ l_{13}+1,l_{32}+1\\p+1
		\end{matrix}} \bigg\rangle_u & p<0 \end{cases}
	\end{align}
	where, 
	\begin{align}
	\bar{g}_1^{322}=l_{12} \hspace{1cm} \bar{h}_1^{322}=|p|+l_{12} \hspace{1cm} \bar{h}_2^{322}=l_{12}
	\end{align}
	
\end{itemize}
\end{enumerate}
Action of all other invariant operators can be calculated from the action of the invariant operators (1), (2) and (4) above. We will illustrate this for the operators $a_1\cdot b_2$, $\epsilon b_3b_2a^\dagger_2$ and $\epsilon a_1 a_2 a_3$.
\begin{enumerate}
	\item $a_1\cdot b_2$
	\begin{align}
	\Big[b_2\cdot b^\dagger_3,a_1\cdot a^\dagger_2\Big]=-\tilde{N}_2 (a^\dagger_2 \cdot b^\dagger_3)(a_1\cdot b_2)
	\end{align}
From eqns (\ref{aafinal}) and (\ref{bbfinal}), we know 
{\footnotesize
\begin{align}
&\Big[b_2\cdot b^\dagger_3,a_1\cdot a^\dagger_2\Big] |l,p\rangle_u = \Big(\bar{d}_1^{21}(l,p)\bar{e}_1^{32}(l_{13}-1,l_{23}+1)+\bar{d}_2^{21}(l,p)\bar{e}_2^{32}(l_{21}+1,l_{12}-1,l_{31}-1,l_{32}+1)\nonumber\\
&-\bar{e}_1^{32}(l,p)\bar{d}_1^{21}(l_{13}+1,l_{12}-1)-\bar{e}_2^{32}(l,p)\bar{d}_2^{21}(l_{23}+1,l_{32}-1,l_{23}-1,l_{31}+1)\Big)|l_{12}-1,l_{23}+1\rangle_u \nonumber\\
& \Big(\bar{d}_1^{21}(l,p)\bar{e}_2^{32}(l_{13}-1,l_{23}+1\Big)-\bar{e}_2^{32}(l,p)\bar{d}_1^{21}(l_{23}+1,l_{32}-1,l_{21}-1,l_{31}+1\Big) \Bigg|{\scriptsize\begin{matrix}
l_{21}-1\\l_{23}+2,l_{32}-1\\l_{31}+1,l_{13}-1
\end{matrix}}\Bigg\rangle_u \nonumber\\
&\Big(\bar{d}_2^{21}(l,p)\bar{e}_1^{32}(l_{21}+1,l_{12}-1,l_{31}-1,l_{32}+1)-\bar{e}_1^{32}(l,p)\bar{d}_2^{21}(l_{13}+1,l_{12}-1)\Big)\Bigg|{\scriptsize\begin{matrix} l_{12}-2,l_{21}+1\\l_{31}-1,l_{13}+1\\l_{32}+1\end{matrix}}\Bigg\rangle_u
	\end{align}}
 Substituting values for $d_1^{21},d_2^{21},e_1^{32},e_2^{32}$ and simplifying gives (\ref{finalab}).
 \item $\epsilon b_3b_2a^\dagger_2$
 \begin{align}
 \Big[b_3\cdot b^\dagger_1,\epsilon a^\dagger_3b_2a^\dagger_2\Big]=-\tilde{N}_3(a^\dagger_3\cdot b^\dagger_1)(\epsilon b_3b_2a^\dagger_2)
 \end{align}
 Since we know the action of the left hand side of above equation on $|l,p\rangle$ from (\ref{bbfinal}), (\ref{adbadp+}) and (\ref{ad3b2ad2p-}), we get,
 {\footnotesize
 \begin{align}
 \hspace{-.5cm}
% &p>0:\nonumber\\
(\epsilon b_3b_2a^\dagger_2)|l,p\rangle_u=\begin{cases}\bar{i}_1^{322}\Bigg|{\scriptsize\begin{matrix}
 l_{12}-1,l_{21}+1\\l_{23}-1,l_{31}-1\\p+1\end{matrix}}\Bigg\rangle_u+\bar{i}_2^{322}\Bigg|{\scriptsize\begin{matrix}l_{13}-1,l_{32}-1\\p+1\end{matrix}}\Bigg\rangle_u
 &p\geq0\\
\bar{j}_1^{322}\Bigg|l_{21}+1,p+1\Bigg\rangle_u+\bar{j}_2^{322}\Bigg|{\scriptsize\begin{matrix}
l_{13}-1,l_{31}+1\\l_{32}-1,l_{23}+1\\l_{12}+1,p+1
\end{matrix}}\Bigg\rangle_u+\bar{j}_3^{322} \Bigg|{\scriptsize \begin{matrix}
l_{12}-1,l_{21}+2\\l_{23}-1,l_{32}+1,p+1\\l_{13}+1,l_{31}-1
\end{matrix}}\Bigg\rangle _u&p<0 \end{cases}
\label{bbad}
 \end{align}}
 where, 
{ \footnotesize
 \begin{align}
 &\bar{i}_1^{322}= -(l_{13}+l_{31}+l_{23}+l_{32}+|p|+2)\Big(\bar{g}_1^{322}(l,p)\bar{e}_1^{12}(l_{12}-1,p+1)-\bar{e}_1^{12}(l,p)\bar{g}_1^{322}(l_{21}+1,l_{23}-1)\Big)\nonumber\\
 &=\frac{-l_{12}l_{23}l_{31}}{(n_3+m_3+|p|+1)}\nonumber\\
 &\bar{i}_2^{322}=-(l_{13}+l_{31}+l_{23}+l_{32}+|p|+2)\Big(\bar{g}_1^{322}(l,p)\bar{e}_2^{12}(l_{12}-1,p+1)\nonumber\\
 &-\bar{e}_2^{12}(l,p)\bar{g}_1^{322}(l_{32}-1,l_{31}+1,l_{13}-1,l_{31}+1)\Big) =-\frac{(n_3+m_3+|p|+2)}{(n_3+m_3+|p|+1)}l_{13}l_{32} \nonumber\\
 &\bar{j}_1^{322}= -(l_{13}+l_{31}+l_{23}+l_{32}+|p|+2)\Big(\bar{h}_1^{322}(l,p)\bar{e}_1^{13}(l_{23}+1,l_{31}+1,p+1)\nonumber\\
 &+\bar{h}_2^{322}(l,p)\bar{e}_2^{12}(l_{12}-1,l_{21}+1,l_{32}+1,l_{13}+1,p+1)-\bar{e}_1^{13}(l,p)h_1^{322}(l_{21}+1,l_{23}-1)\nonumber\\
 &-\bar{e}_2^{13}(l,p)\bar{h}_2^{322}(l_{13}-1,l_{31}+1,l_{32}-1,l_{12}+1)\Big)\nonumber\\
 &=l_{12}(l_{13}+1)(l_{32}+1)-\frac{(n_3+m_3+|p|+2)}{(n_3+m_3+|p|+1)}l_{13}l_{32}(l_{12}+1)\nonumber\\
 &-\frac{(|p|+l_{12})(n_3+m_3-l_{23}+|p|+1)(n_3+m_3-l_{31}+|p|+1)}{(n_3+m_3+|p|+1)}\nonumber\\
 &\bar{j}_2^{322}=-(l_{13}+l_{31}+l_{23}+l_{32}+|p|+2)\Big(\bar{h}_1^{322}(l,p)\bar{e}_2^{13}(l_{23}+1,l_{31}+1,p+1)\nonumber\\
% \nonumber\\
 &-\bar{e}_2^{13}(l,p)\bar{h}_1^{322}(l_{13}-1,l_{31}+1,l_{32}-1,l_{13}+1)\Big)=- \frac{(n_3+m_3+2|p|+l_{12}+2)}{(n_3+m_3+|p|+1)}l_{13}l_{32} \nonumber\\
 &\bar{j}_3^{322}=-(l_{13}+l_{31}+l_{23}+l_{32}+|p|+2)\Big(\bar{h}_2^{322}(l,p)\bar{e}_1^{13}(l_{12}-1,l_{21}+1,l_{32}+1,l_{13}+1,p+1)\nonumber\\
 &-\bar{e}_1^{13}(l,p)\bar{h}_2^{322}(l_{21}+1,l_{23}-1\Big) =-\frac{(l_{12}l_{23}l_{31})}{n_3+m_3+|p|+1}
 \label{ijbar}
 \end{align}
}
 \item $\epsilon a_3a_2b^\dagger_2$

 Since, this operator can be obtained from $\epsilon b_3b_2a^\dagger_2$ by $a\leftrightarrow b$, the action of this operator can be obtained by the symmetry  operation 1 discussed in section (\ref{su3}). 
  {\footnotesize
  \begin{align}
 % &p>0:\nonumber\\
 (\epsilon a_3a_2b^\dagger_2)|l,p\rangle_u=\begin{cases}\bar{k}_1^{322}\Bigg|l_{12}+1,p-1\Bigg\rangle_u+\bar{k}_2^{322}\Bigg|{\scriptsize\begin{matrix}
 	l_{31}-1,l_{13}+1\\l_{23}-1,l_{32}+1\\l_{21}+1,p-1
 	\end{matrix}}\Bigg\rangle_u+\bar{k}_3^{322} \Bigg|{\scriptsize \begin{matrix}
 	l_{21}-1,l_{12}+2\\l_{32}-1,l_{23}+1,p-1\\l_{31}+1,l_{13}-1
 	\end{matrix}}\Bigg\rangle_u
 &p>0\\\bar{l}_1^{322}\Bigg|{\scriptsize \begin{matrix}
 l_{21}-1,l_{12}+1\\l_{32}-1,l_{13}-1\\p-1\end{matrix}}\Bigg\rangle_u+\bar{l}_2^{322}\Bigg|{\scriptsize \begin{matrix}l_{31}-1,l_{23}-1\\p-1\end{matrix}}\Bigg\rangle_u
  &p\leq0 \end{cases}
 \label{aabd}
 \end{align}}
 	\begin{align}
 	& \bar{l}_1^{322}=\bar{i}_1^{322}|_{l_{ij}\leftrightarrow l_{ji}, p+1\leftrightarrow p-1}\hspace{1cm}
 	\bar{l}_2^{322}= \bar{i}_2^{322}|_{l_{ij}\leftrightarrow l_{ji}, p+1\leftrightarrow p-1}\nonumber\\
 	&\bar{k}_1^{322}= \bar{j}_1^{322}|_{l_{ij}\leftrightarrow l_{ji}, p+1\leftrightarrow p-1}\hspace{1cm}\bar{k}_2^{322}=\bar{j}_2^{322}|_{l_{ij}\leftrightarrow l_{ji}, p+1\leftrightarrow p-1}\nonumber\\
 	&\bar{k}_3^{322}=\bar{j}_3^{322}|_{l_{ij}\leftrightarrow l_{ji}, p+1\leftrightarrow p-1}
 	\end{align}
 	\item $(\epsilon a_1 a_2 a_3)$
 	\begin{align}
 	\Big[a_1\cdot b_2, \epsilon a_3a_2b^\dagger_2 \Big]=(\epsilon a_3 a_2 a_1)
 	\end{align}
 	The action of the LHS of the above equation on $|l,p\rangle_u$ is easily obtained using equations (\ref{aabd}), (\ref{finalab}) thereby giving $(\epsilon a_3 a_2 a_1)|l,p\rangle_u$.
\end{enumerate}
\section{SU(3): norm }
Since we know the action of $a_i\cdot b_j$ on the $|l,p\rangle_u$, the norm $S(l,p)$ of the unnormalised state can be calculated using the following recursion relation:
\begin{align}
&S(l,p)=
{}_u\langle l,p|l,p\rangle_u={}_u\langle l_{12}-1|(a_1\cdot b_2)|l,p\rangle_u={\bar f}_1^{12}(l_{12}){}_u\langle l_{12}-1| l_{12}-1\rangle_u\nonumber\\
&=\bar{f}_1^{12}(l_{12}) S(l_{12}-1)
\end{align}
To calculate $S(l,p)$, we further need $(\epsilon^{~}_{\alpha\beta\gamma} a^{~}_{1,\alpha}a^{~}_{2,\beta}a^{~}_{3,\gamma})|l=0,p\rangle$ which is computed as : 
\begin{align}
&(\epsilon^{~}_{rst} a^{~}_{1,r}a^{~}_{2,s}a^{~}_{3,t})(\epsilon_{rst} a^{\dagger}_{1,r}a^{\dagger}_{2,s}a^{\dagger}_{3,t})|l=0,p-1\rangle=(a_1\cdot a^\dagger_1)(a_2\cdot a^\dagger_2)(a_3\cdot a^\dagger_3)|l=0,p-1\rangle\nonumber\\
&=(|p|+2)^3|l=0,p-1\rangle
\end{align}
When $p<0$, a similar calculation goes through and gives the same result.
The norm is given by : 
%\begin{widetext}
	\begin{align}
	S(l,p)=&F_{12}(l,p)~F_{21}(l,p)\Big|_{l_{12}=0}~F_{13}(l,p)\Big|_{l_{12}=l_{21}=0}
	%\nonumber\\
	F_{31}(l,p)\Big|_{l_{12}=l_{21}=l_{13}=0}~F_{23}(l,p)\Big|_{l_{12}=l_{21}=l_{13}=l_{31}=0}
\nonumber\\
&	F_{32}(l,p)\Big|_{l_{12}=l_{21}=l_{13}=l_{31}=l_{23}=0}
	~~~\frac{[(p+2)!]^3}{2^3}\nonumber\\
		\label{norm}
	\end{align}
	%%Above, 
	where
	\begin{align}
	F_{12}(l,p)=&\bar{f}_1^{12}(l_{12})\bar{f}_1^{12}(l_{12}-1)\cdots
	%\nonumber\\
	\bar{f}_1^{12}(l_{12}=1) 
	\end{align}
	and similar expressions for $F_{21}(l,p),F_{13}(l,p),F_{31}(l,p),F_{23}(l,p),F_{32}(l,p)$.
%\end{widetext}

\end{document}